\documentclass[sigconf,authorversion,nonacm]{acmart}
\usepackage{amsmath,nccmath,tabularx} 
\usepackage{algorithm}
\usepackage{algorithmic}
\usepackage{graphicx}
\graphicspath{{images/}}
\usepackage{wrapfig,lipsum}
\usepackage{tabularx}
\usepackage{subcaption}
\usepackage{multirow}
\usepackage{hyperref}
\usepackage{makecell}
\usepackage{fixmath}
\usepackage{enumitem}

\AtBeginDocument{%
  \providecommand\BibTeX{{%
    \normalfont B\kern-0.5em{\scshape i\kern-0.25em b}\kern-0.8em\TeX}}}

\copyrightyear{2021} 
\acmYear{2021} 
\setcopyright{acmlicensed}\acmConference[]{}{}{}
\acmBooktitle{Arxiv Preprint}
\acmPrice{}
\acmDOI{}

\begin{document}
\title{Fast Passage Re-ranking with Contextualized Exact Term Matching and Efficient Passage Expansion}


\author{Shengyao Zhuang}
\affiliation{%
	\institution{The University of Queensland}
	\streetaddress{4072 St Lucia}
	\city{Brisbane}
	\state{QLD}
	\country{Australia}}
\email{s.zhuang@uq.edu.au}

\author{Guido Zuccon}
\affiliation{%
	\institution{The University of Queensland}
	\streetaddress{4072 St Lucia}
	\city{Brisbane}
	\state{QLD}
	\country{Australia}}
\email{g.zuccon@uq.edu.au}


\begin{abstract}

BERT-based information retrieval models are expensive, in both time (query latency) and computational resources (energy, hardware cost), making many of these models impractical especially under resource constraints. The reliance on a query encoder that only performs tokenization and on the pre-processing of passage representations at indexing, has allowed the recently proposed TILDE method to overcome the high query latency issue typical of BERT-based models.  This however is at the expense of a lower effectiveness compared to other BERT-based re-rankers and dense retrievers. In addition, the original TILDE method is characterised by indexes with a very high memory footprint, as it expands each passage into the size of the BERT vocabulary.

	In this paper, we propose TILDEv2, a new model that stems from the original TILDE but that addresses its limitations.
	TILDEv2 relies on contextualized exact term matching with expanded passages. This requires to only store in the index the score of tokens that appear in the expanded passages (rather than all the vocabulary), thus producing indexes that are 99\% smaller than those of TILDE. This matching mechanism also improves ranking effectiveness by 24\%, without adding to the query latency. This makes TILDEv2 the state-of-the-art passage re-ranking method for CPU-only environments, capable of maintaining query latency below 100ms on commodity hardware.
	
	One potential drawback of TILDEv2, compared to the original TILDE, is the extra passage expansion process required at indexing. This is an expensive process if performed using current passage expansion methods. However, we address this by adapting the original TILDE model to serve as a passage expansion method. Compared to current expansion methods, our proposed method reduces the passage expansion time by 98\% with only less than 1\% effectiveness loss on the MS MARCO passage ranking dataset (and even improvements on other datasets). We further show that our expansion approach generalises to other ranking methods that rely on expansion. 
\end{abstract}




\maketitle

\section{Introduction} \label{intro}
Passage ranking is a core task for many web search and information retrieval applications. Traditional passage retrieval methods, such as BM25, rely on exact lexical matching signals and use frequency-based term importance estimation to calculate the matching score between queries and passages. This bag-of-words (BOW) mechanism however limits the capability of these methods of retrieving passages that are semantically relevant but have low or zero query term frequency: the well-known \textit{vocabulary mismatch problem}~\cite{furnas1987vocabulary}.

Neural retrieval methods aim to address this limitation. Recent advances in neural rankers have seen the introduction of pre-trained deep language models (LMs) that are then fine-tuned on passage ranking tasks. These methods leverage the contextualized representation produced by deep LMs~\cite{peters2018deep}, such as BERT~\cite{devlin2019bert}, to estimate the semantic matching score between passages and queries. For example, monoBERT~\cite{nogueira2019passage} takes query-passage pairs as the input of BERT and the matching scores are estimated on the contextualized CLS token representation. Many empirical results obtained with monoBERT and its variants~\cite{dai2019deeper,gao2021lce,nogueira2019multi,nogueira2020document} have demonstrated that these deep LMs based rankers achieve substantially better effectiveness than BOW methods on passage ranking tasks. However, this effectiveness gain does not come for free. The query latency cost of this type of neural rankers is several orders of magnitude larger than that of BOW methods~\cite{nogueira2019multi}. In addition, GPUs are required, in place of more economical CPUs, not only for the offline training of the rankers, but also for the online (i.e. at query time) encoding of the contextualized representations. This hinders the practical adoption of these powerful rankers on small, GPU-free devices, such as mobile phones or embedded systems, or limits the number of passages that can be considered for re-ranking within a reasonable amount of time to guarantee real-time responses~\cite{hofstatter2020interpretable,tilde2021zhuang}.

To address the high query latency issue of these BERT-based re-rankers and to allow for the use of CPU-based systems in place of GPU-based ones, the recently proposed TILDE method~\cite{tilde2021zhuang} proposes to only use the BERT tokenizer to encode query representations at query time (online), while use BERT to pre-compute contextualized token importance scores over the BERT vocabulary for each passage in the index at indexing time (offline). Since no BERT inference is required at query time, TILDE achieves impressive re-ranking speed and GPUs are not required for inference. However, despite being very efficient, TILDE is much less effective than state-of-the-art BERT-based re-rankers and dense retrievers: TILDE trades off query representation quality, and thus effectiveness, for querying efficiency. On the other hand, because TILDE expands all passages in the collection to the size of the BERT vocabulary, it also has the drawback of a large index (and thus associated memory requirements): each passage, in fact, has a posting for every term in the BERT vocabulary.

In this paper, we propose changes to the TILDE method that tackle the current drawbacks of TILDE, while maintaining its efficiency. The result is a method, referred to as TILDEv2, that is highly efficient in both query latency (maintaining TILDE's original low latency) and index size (reducing the original TILDE index by up to 99\%), and showcases effectiveness at par to state-of-the-art BERT-based methods (improving over TILDE by up to 24\% on MS MARCO). This makes TILDEv2 to be production-ready for applications with limited computational power (e.g. no GPUs, start-ups). Specifically, we modify TILDE in the following aspects:

\begin{itemize}
	\item \textbf{Exact Term Matching.} The query likelihood matching originally employed in TILDE, expands passages into the BERT vocabulary size, resulting in large indexes. To overcome this issue, we follow the recent paradigm of estimating relevance scores with contextualized exact term matching~\cite{gao2021coil,lin2021few,mallia2021learning}. This allows the model to index tokens only present in the passage, thus reducing the index size. In addition to this, we replace the query likelihood loss function, with the \textit{Noise-contrastive estimation} (NCE) loss~\cite{gutmann2010noise} that allows to better leverage negative training samples. This loss function has been shown effective for recent dense retrievers~\cite{qu2021rocketqa,karpukhin2020dense,lin2021batch,gao2021coil} and BERT-based re-rankers~\cite{gao2021lce}.
	\item \textbf{Passage Expansion.} To overcome the vocabulary mismatch problem that affects exact term matching methods, we use passage expansion to expand the original passage collection. Passages in the collection are expanded  using deep LMs with a limited number of tokens. This requires TILDEv2 to only index a few extra tokens in addition to those in the original passages.
\end{itemize}

%


Compared to the original TILDE, the only drawback introduced by TILDEv2 is an extra passage expansion process to be executed at indexing time. This passage expansion process is typical of methods that exploit contextualized exact term matching~\cite{gao2021coil,lin2021few,mallia2021learning}. For passage expansion, these previous methods rely on docT5query~\cite{nogueira2019doc2query} to generate related tokens that are appended to the original passage. 
However, docT5query is a T5-based~\cite{2020t5} sequence-to-sequence generative model, which is very expansive for inference: it requires 320 hours on a preemptible TPU\footnote{Estimated based on generating 40 ``expansion queries'' per passage.} to expand the whole MS MARCO passage collection. This process becomes then expansive, and often infeasible, for large-scale information retrieval applications such as web search, or for small organisations such as start-ups. In TILDEv2 instead, we introduce a new way of performing this passage expansion process by replacing docT5query with the original TILDE model (which is then used for passage expansion, but not for retrieval). Empirical evaluation demonstrates that the proposed passage expansion method requires only a fraction of the time of the previous expansion method (45 times faster than docT5query), with only less than 1\% effectiveness loss, if any. In addition, we also show that our passage expansion method is generally applicable to other retrieval methods such as uniCOIL~\cite{lin2021few}.


\begin{figure*}[t]
	\includegraphics[width=\linewidth]{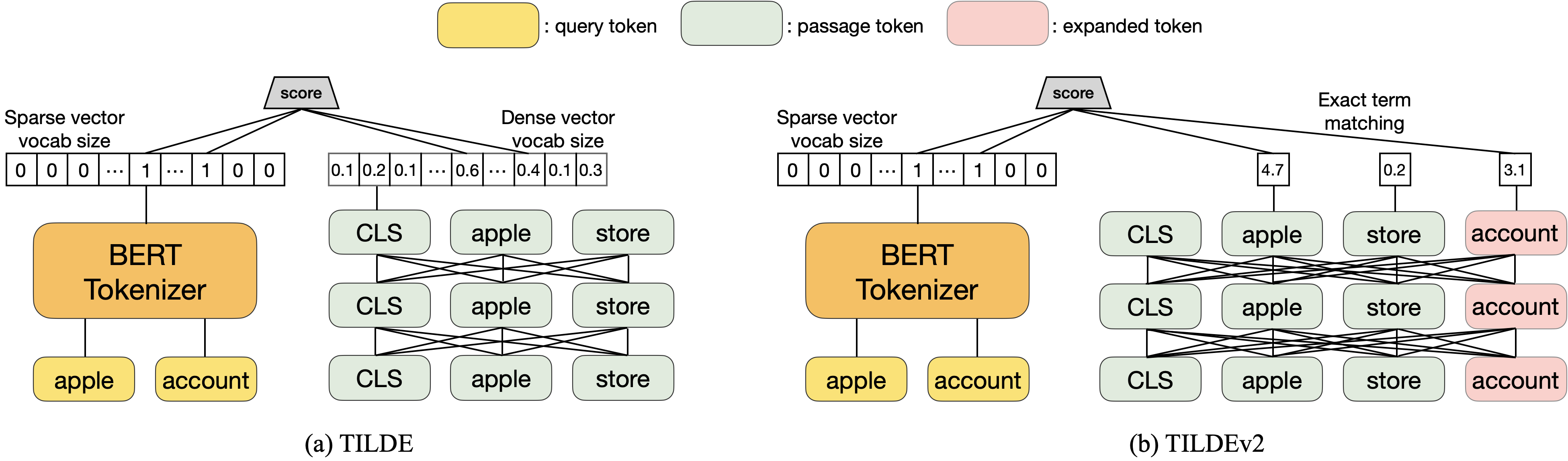}
	\caption{Model architectures. Left: the original TILDE model. Right: Our TILDEv2 model.}
	\label{fig:model}
\end{figure*}

\section{Related work} \label{related}

Transformer-based~\cite{vaswani2017attention} pre-trained deep LMs, such as BERT~\cite{devlin2019bert}, have been shown to provide rich contextualized information, delivering high effectiveness in many NLP~\cite{brown2020language,yang2019xlnet,raffel2020exploring} and downstream retrieval and ranking tasks~\cite{lin2020pretrained}.
\citeauthor{nogueira2019passage} were the first to directly fine-tuned BERT for passage ranking, achieving a large performance leap over BOW methods~\cite{nogueira2019passage}. \citeauthor{gao2021lce} further showed that BERT trained with localized NCE loss achieves better effectiveness for document re-ranking~\cite{gao2021lce}. Beyond BERT, other generative pre-trained LMs such as GPT~\cite{brown2020language}, BART~\cite{lewis2020bart} and T5~\cite{raffel2020exploring} also have shown promising results in text re-ranking with query likelihood~\cite{dos2020beyond,zhuang2021deep,lesota2021modern}. 

The biggest shortage of BERT-based rankers is their high query latency: several expensive BERT inferences are needed for ranking passages from a candidate pool because at query time BERT requires as input individual query-passage pairs.
This makes BERT-based rankers feasible only for re-ranking, and with small rank cut-offs.


Several works have attempted to address the high query latency of BERT-based re-rankers. One direction is modifying the BERT encoder. 
\citeauthor{hofstatter2020interpretable}~\cite{hofstatter2020interpretable} proposed the Transformer Kernel (TK): Instead of using the full-size BERT, TK uses a limited number of transformer layers to pre-compute contextual representations of the passages' tokens, requiring only a small amount of computation to produce the query representation at query time. The encoded query tokens and document tokens are then used to compute a feature matrix with a kernel operation. 
\citeauthor{macavaney2020expansion}~\cite{macavaney2020expansion} proposed EPIC, where query and passages are encoded independently: this allows to pre-compute the passage representation at indexing time. 
The original TILDE method~\cite{tilde2021zhuang}, which we build on top of in this paper, takes this idea to the extreme: TILDE only uses the BERT tokenizer to encode the query representation at query time. 
Since no transformer layer is involved at query time, TILDE can be ran efficiently on a CPU-only environment. A drawback of both TILDE and EPIC is the large amount of memory required to store the passage representations (index).

An alternative direction is reconsidering the representation. This is the approach followed by BERT-based dense retrievers (DRs)~\cite{zhan2020repbert,xiong2020approximate,khattab2020colbert,qu2021rocketqa,karpukhin2020dense,gao2021clear}.
Similar to EPIC
, DRs also compute passage representations at indexing time, rather than at querying, requiring then at query time only one BERT inference for encoding the query. Passages are ranked using the similarity between the query representation and the passage representations: when powerful GPUs are used, DRs can achieve similar query latency as traditional BOW methods based on inverted index (e.g., BM25). 

A third direction is represented by methods based on the supervised construction of an inverted index and the use of contextualized exact term matching. 
DeepCT~\cite{dai2019context}, for example, uses BERT to estimate the contextualized term importance weight for terms in a passage. Then, the learnt term weights are stored in a standard inverted index. Hence, retrieval can be performed using any BOW exact term matching method, such as BM25. DeepCT's effectiveness, however, is limited by the exact term matching mechanism as it can only estimate weights for terms that appear in the passage. This drawback is solved in the recently proposed DeepImpact~\cite{mallia2021learning}: DeepImpact first uses docT5query~\cite{nogueira2019doc2query} to generate terms to expand the original passage. Contextualized term weights are then assigned to these new terms, and matching can be executed. 
COIL~\cite{gao2021coil} uses a different contextualized exact term matching architecture. Instead of storing scalar term weights, COIL learns and stores contextualized token representations into the inverted lists, which are then used to perform exact term matching at query time. A subsequent work by \citeauthor{lin2021few} introduced a variation of COIL, named uniCOIL~\cite{lin2021few}, which combines the idea of passage expansion from DeepImpact with the COIL architecture, so as to learn scalar weights that can be stored in the standard inverted index. Our proposed TILDEv2 builds upon this prior work by introducing contextualized exact term matching coupled with passage expansion into TILDE for the passage re-ranking task.


\section{Method} \label{method}
The proposed TILDEv2 addresses the limitations of the current TILDE model by integrating and expanding upon a number of recent advances in BERT-based rankers. Next, we discuss the key components of our TILDEv2 and its similarities and differences compared to the original TILDE.

\subsection{Tokenizer-based query encoder}\label{sec:tokenzier}
Unlike other BERT-based re-rankers and DRs, at query time TILDE only uses the BERT tokenizer to encode the query into a sparse vector representation: this represents one of the main innovations of TILDE. This tokenizer-based query encoder is very simple but also very efficient: it is a lookup table without any model parameters, thus eliminating the need for costly inferences that require GPU computation. In our experiments, it takes only less than 1 ms to encode the query with TILDE. 

In order to achieve maximum re-ranking speed, our TILDEv2 inherits the simple query encoder which is at the basis of TILDE. As Figure~\ref{fig:model} illustrates, both TILDE and TILDEv2 use the BERT tokenizer to encode the query. The encoded query representation is a sparse vector of dimension equal to the BERT vocabulary size, in which each element in the vector is the frequency of that token in the query. For instance, for the query `apple account' (Figure~\ref{fig:model}), the token \textit{`apple'} appears once in the query and its token id in the BERT vocabulary is 6207; hence, the value of the 6207th element in the query vector is 1. These non-zero elements are used to compute the matching score between the query and a passage (see next). Since queries in web search are often short~\cite{silverstein1999analysis}, the vectors are sparse (only few elements are non-zero), and the matching operation only requires to consider the non-zero elements, being thus very efficient.

\subsection{Re-ranking with contextualized exact term matching}

The biggest difference between TILDE and TILDEv2 is the query-passage relevance matching mechanism. TILDE follows the query likelihood paradigm~\cite{ponte1998language}, where the probability of a passage being relevant to a query is estimated by the likelihood of generating the query text given the passage text. As Figure~\ref{fig:model}(a) illustrates, TILDE outputs the query token probabilities over the BERT vocabulary (presented as a vocabulary size dense vector)  using a projection layer on top of the BERT's [CLS] token. Then, it assumes the query tokens are independent and it computes the query likelihood by summing the log probabilities of the query tokens in the dense vector. This relevance matching mechanism forces TILDE to compute and store query token likelihoods for all tokens in the BERT vocabulary at indexing time, resulting in a very large index. As an example, the TILDE index size for the MS MARCO passage collection is more than 500 GiB, compared to the standard Lucene index which is only 2.8GiB: TILDE index size is then often not practical, especially for the systems it targets (those with low computational power, i.e. no GPUs, like mobiles and embedded systems, which are often also characterised by limited disk space).

To overcome this issue, TILDEv2 abandons the use of the query likelihood matching mechanism. Inspired by recent advances in contextualized term weighting, instead, in TILDEv2 we use BERT to output a scalar importance weight for all tokens in the passage and perform exact term matching between query and passage tokens. Figure~\ref{fig:model}(b) illustrates the matching mechanism used in TILDEv2.

Specifically, we use BERT to output the contextualized token embeddings for all passage tokens, and then we feed them into a projection layer to downsize the embeddings to dimension 1:
\begin{equation}
v_{i}^{p} =ReLU\left( \mathbold{W}_{proj}^{1\times n} \mathbold{BERT} (p_i) + b \right)
\end{equation}
where $p_i$ is the $i$-th token in the passage and $\mathbold{W}_{proj}^{1\times n}$ is the projection matrix that maps the BERT's $n$ dimensional word embedding of $p_i$ into a scalar; $b$ is the learnable bias parameter of the projection layer. The scalar output by the projection layer is then passed to a ReLU operation to obtain the final contextualized term importance weight $v_{i}^{p}$ for $p_i$. The ReLU operation masks out all the negative scalars to zero, thus forcing all term weights to be positive. This operation has also been used in previous work~\cite{lin2021few,mallia2021learning}.

We now define the exact term matching scoring function for computing the passage relevance scores given all the query tokens' frequencies encoded by the BERT tokenizer:
\begin{equation}
S(q, p) = \sum_{q_i \in q} \max_{q_i=p_j} \left( c(q_i) \times v_{j}^{p}  \right)
\end{equation}
where $c(q_i)$ is the count of the $i$-th unique query token given by the BERT tokenizer query encoder described in section~\ref{sec:tokenzier}. The relevance score of each query-passage pair is the sum of the contextualized term weights provided by each query token that appears in the passage. If a query token appears more than once in a passage, then its score is equal to the highest contextualized term weight ($max(.)$) for that token in the passage.
 With this matching mechanism, TILDEv2 only needs to pre-compute and store the tokens that appear in the passage along with the max contextualized term weight. Compared to TILDE, which needs to store the likelihood value of all tokens in the BERT vocabulary, the index size of TILDEv2 is therefore two orders of magnitude smaller (we provide more details on this aspect in section~\ref{sec:expansion_impact}). 

Finally, following previous work~\cite{gao2021lce,gao2021coil,karpukhin2020dense,lin2021batch,qu2021rocketqa}, we train our TILDEv2 with the NCE loss function~\cite{gutmann2010noise} with the negative passage set $l$ created by randomly sampling passages from the top 1,000 results obtained by BM25 for the query:
\begin{equation}
\mathcal{L} = -\log \frac{\exp({S(q,p^+)})}{\exp({S(q,p^+)}) + \sum_{p^-\in l}\exp({S(q,p^-)})}
\end{equation}
 More training details are discussed in section~\ref{sec:train}
 
\begin{algorithm}[t]
	\caption{ Passage expansion with the original TILDE model.}
	\begin{algorithmic}[1]
		\STATE {\bf Input}: $TILDE$, stopword list $ stop\_list$, passage $p$,  Threshold $m$
		\STATE $T = TILDE(p)$\hfill// get token likelihood distribution $T$.
		\STATE $sort(T)$ \hfill// sort by descending order of the likelihoods.
		\FOR{$t \in T[ :m] $}
		\IF {$t \notin p$ AND $t \notin stop\_list$}
		\STATE $p.append(t)$
		\ENDIF
		\ENDFOR
	\end{algorithmic}
	\label{algo:expansion}
\end{algorithm} 

\subsection{Passage expansion}\label{sec:expansion}
Similar to traditional BOW methods, TILDEv2 can only match those query terms that appear in the passage; thus, if no other matching mechanism is put in place, its effectiveness is limited by the vocabulary mismatch problem. In order to reduce the impact of this problem, following recent advances in exact term matching models~\cite{mallia2021learning,lin2021few}, we use the technique of passage expansion to expand each passage in the collection at indexing time. Passage expansion appends semantically related and potentially relevant terms at the end of a passage, in the bid to increase the likelihood of retrieving the passage for queries containing those expanded terms and for which the passage is relevant.

Existing approaches use docT5query~\cite{nogueira2019doc2query} to perform passage expansion~\cite{mallia2021learning,lin2021few}. docT5query is a T5-based~\cite{raffel2020exploring} sequence-to-sequence generative language model, which can only generate one token at a time. Thus, multiple inferences from docT5query are needed to obtain several tokens for passage expansion. Provided that T5 is a large transformer model, passage expansion with docT5query requires a large amount of computational resources. According to the statistic provided by the docT5query authors~\cite{nogueira2019doc2query}, sampling 40 queries\footnote{This is the common number used in previous work.} per passage for each of the $\approx$8.8 million passages in the MS MARCO collection requires $\approx$320 hours on a single TPU, and $\approx$5,000 hours are required for expanding the MS MARCO v2's 138.3 million passages\footnote{https://microsoft.github.io/msmarco/TREC-Deep-Learning.html}. For large-scale information retrieval applications such as web search, this is a very expensive process.

For TILDEv2 we take a different approach to passage expansion: 
we adapt the original TILDE method to perform the passage expansion. This idea is based on the observation that TILDE actually outputs a query token likelihood distribution over the vocabulary. This distribution can be considered as an estimation of term importance given the passage context. In addition, unlike docT5query, the original TILDE model assumes query terms are independent, so that it only needs a single inference step to get the distribution output for all tokens. 

 The main algorithm that exploits the original TILDE model for passage expansion is described in Algorithm~\ref{algo:expansion}. For a given passage $p$, we use TILDE to get the likelihood distribution $T$. Note, each element in $T$ is a token-likelihood pair. We then sort all the tokens in $T$ in descending order according to their corresponding likelihoods. Next, for each top-$m$ token $t$ in the sorted list $T$, if $t$ is not in the original passage $p$ and it is not in a pre-defined stopword list, we then append it to the original passage. We do this expansion for all passages in the collection. In our experiments, passage expansion with TILDE can expand the whole MS MARCO passage collection in 7.3 hours on a single GPU. More details are provided in section~\ref{sec:expansion_impact}. Note, the use of TILDE for passage expansion was not present in the original work of Zhuang and Zuccon~\cite{tilde2021zhuang}, and thus is a novel contribution of our work.

\section{Experimental settings} \label{experiments}
Next, we describe the experiment settings  we use to investigate the performance of our TILDEv2, to compare it to current, relevant methods in the literature, and to answer the research questions:

 \begin{itemize}
	\item \textbf{RQ1}: Which matching mechanism is more effective and more efficient: the query likelihood matching used in the original TILDE, or the contextualized exact term matching used in TILDEv2?
	\item \textbf{RQ2}: How does TILDEv2 compare to current methods for passage ranking in terms of effectiveness and efficiency?
	\item \textbf{RQ3}: How does the effectiveness-efficiency trade-off of TILDEv2, allowed by the setting of the rank cut-off parameter, compare to that of the BERT re-ranker? 
	\item \textbf{RQ4}: How effective and how efficient is our passage expansion based on TILDE, compared to the current  state-of-the-art method (docT5query)?
\end{itemize}

\subsection{Datasets and evaluation metrics}
We experiment with three commonly used publicly available large-scale passage ranking datasets: MS MARCO~\cite{nguyen2016ms}, TREC Deep Learning 2019~\cite{craswell2020overview} and TREC Deep Learning 2020~\cite{craswell2021overview} (DL2019, DL2020). These datasets share the same set of passages, the MS MARCO passage corpus\footnote{https://github.com/microsoft/MSMARCO-Passage-Ranking}, which consists of approximately 8.8 million passages (average length: 73.1 terms) crawled by the Bing search engine, but differ in terms of queries (and relevance assessments). 

The MS MARCO dataset provides approximately 1 million queries. Queries are split into train, dev, and eval sets. Each query is associated with shallowly annotated judgments, where on average only one passage is marked as relevant and no irrelevant passages are identified. Following standard practice from the dataset instructions, we use queries along with their relevance judgments in the train set to train our model; while we evaluate the model on the dev set. Evaluation is performed with respect to the official evaluation measure MRR@10.

 Unlike MS MARCO, the TREC DL2019 and DL2020 datasets provide small query sets (43 for DL 2019, 54 for DL2020), with deep judgments on a four-point scale (i.e. graded). Following TREC DL practice, we use nDCG@10 and MAP as evaluation measures, so we can more easily compare our method to past and future work. 
 
 For all evaluation measures, differences between methods are tested for statistical significance using a paired two-tailed t-test with Bonferroni correction.

Along with effectiveness, we also report the query latency achieved within a CPU environment and within a GPU environment. For this, we randomly sampled 200 queries from the dev queries of MS MARCO and issued them one by one to each model, and report the average query latency measured. For the CPU environment, we conducted experiments on a consumer-grade 3.2GHz 6-core Intel Core i7 CPU with 64GB DDR4 memory (2018 Apple Mac Mini). For the GPU environment, we used an NVIDIA Tesla V100 16G GPU.

\subsection{Baselines}

\textbf{BOW retrievers}: We consider the traditional BOW approach BM25 and the commonly used strong BOW baseline docT5query~\cite{nogueira2019doc2query}. docT5query uses BM25 for ranking, but it performs passage expansion using the T5 deep LM as a collection pre-processing step. We also use these two methods as first stage retrievers, on top of which we apply TILDEv2 (and other deep LM re-rankers). For both methods, we use the Anserini~\cite{yang2018anserini} implementation with its default parameter setting.

\textbf{Contextualized exact match}: These methods use deep LMs to assign contextualized term weight and perform exact term matching with an inverted index. We use the recent DeepImpact~\cite{mallia2021learning} and uniCOIL~\cite{lin2021few} methods. At query time, Deepimpact uses the BERT tokenizer to ``clean'' the query tokens. uniCOIL, instead, performs a BERT inference to compute the contextualized term weights for the query tokens. For uniCOIL, we use the GitHub code\footnote{https://github.com/luyug/COIL/tree/main/uniCOIL} provided by the authors to train the model and use Anserini to index the collection. For DeepImpact, we directly use the Anserini implementation. 

\textbf{Dense Retrievers}: We also consider dense retrievers, and specifically RepBERT~\cite{zhan2020repbert} and ANCE~\cite{xiong2020approximate}, as means of very efficient neural methods for retrieval. RepBERT uses BERT to encode the query and the passages and is trained with BM25 hard negatives. ANCE uses RoBERTa~\cite{liu2019roberta}, a more robust version of BERT, as the encoder. For both methods, we use the model checkpoints provided by the authors and the FAISS~\cite{JDH17} Python toolkit to build a dense vector index.

\textbf{BERT-based re-rankers}: We consider two types of BERT-based re-rankers. EPIC~\cite{macavaney2020expansion} is a fast re-ranker that uses BERT to pre-encode passages and at query time it performs a single BERT inference to encode the query. The re-ranking is then performed using similarity matching between the query and passage representations. For EPIC, we use the implementation available in the OpenNIR toolkit~\cite{macavaney:wsdm2020-onir}. The BERT-base/BERT-large re-ranker~\cite{nogueira2019passage} (also known as monoBERT) are strong BERT-based re-ranker baselines. This approach requires that both the query and the passage are jointly provided at query time as inputs to BERT; the output is the matching score. The BERT-large re-ranker differs from the BERT-base from the (larger) number of parameters. We use the model checkpoints made publicly available by Huggingface's model hub~\cite{wolf-etal-2020-transformers} and provided by the NBoost IR platform~\cite{koursaros2019NBoost}.

\textbf{Tokenizer-based re-rankers}: At query time, both the original TILDE~\cite{tilde2021zhuang} and TILDEv2 only use the BERT tokenizer to encode the query. The key difference between the two methods is their matching mechanism. By comparing our TILDEv2 with the original TILDE we can directly measure the impact of our additions in TILDEv2 with respect to effectiveness, query latency and index size.
For the original TILDE model, we use the model checkpoint made available by the authors on the Huggingface model hub.

\subsection{TILDEv2 implementation and training}\label{sec:train}

We implemented TILDEv2 using Pytorch and the Huggingface transformers library~\cite{wolf-etal-2020-transformers}. We used the bert-base-uncased, which has 110M parameters, as BERT model in TILDEv2. The contextualized word embeddings output by BERT have a dimension of 768; they are then projected to scalars (dimension of 1) by a projection layer. As in the original TILDE, we filtered out the same set of stopwords when encoding queries with the BERT tokenizer. 

We trained TILDEv2 for 204,000 update steps with the AdamW optimizer. The learning rate was set to 3e-6 with a linear warm-up schedule. Following previous work~\cite{gao2021coil,gao2021lce,lin2021few} we used both in-batch negatives and hard negatives for the NCE loss function; these were sampled from the BM25 top 1,000 results. More specifically, for each query we sampled 7 hard negatives from BM25 and one positive. We set the batch size to 8, resulting in a total of 63 negatives per query (7 hard negatives + 56 in-batch negatives). The model was trained on a single NVIDIA Tesla V100 16G GPU; training took approximately 10 hours. We used the Python built-in dictionary class (Hashtable) to implement the index, which is used at re-ranking to search the stored contextualized term weights for the tokens in the passages.

\section{Results} \label{results}

%
%

\begin{table*}[t]
	\caption {Effectiveness and efficiency of TILDEv2 and baselines. Statistical significant differences ($p<0.05$) in effectiveness between TILDEv2 and the baselines is reported with subscripts.  The average query latency is measured in milliseconds. The latency of the re-ranking methods includes that of the first stage retrieval. The BOW models, TILDE and TILDEv2 do not run on GPU; while executing the BERT-base/large re-rankers in a CPU environment is infeasible.} \label{tbl-results}
	\centering
		\begin{tabular}{ l | l|ll|ll|cc}
			\toprule
			& \multicolumn{1}{c|}{\bf MS MARCO} & \multicolumn{2}{c|}{\bf TREC DL2019} & \multicolumn{2}{c|}{\bf TREC DL2020} & \multicolumn{2}{c}{\bf Latency (ms)} \\ \toprule
			\bf Method                                & \bf MRR@10                        & \bf nDCG@10 &        \bf MAP         & \bf nDCG@10 &        \bf MAP         & \bf GPU &          \bf CPU           \\ \midrule
			\textbf{(i) BOW retriever}      &                                   &             &                        &             &                        &         &                            \\
			a) BM25                                      & 0.187                             &    0.506    &         0.377          &    0.480    &         0.286         &    n.a.    &           $70$            \\
			b) docT5query (d2q)                            & 0.277                             &    0.648    &         0.463          &    0.616    &         0.408          &    n.a.   &           $75$            \\ \midrule
			\textbf{(i) Contextualized exact match}          &                                   &             &                        &             &                        &         &                            \\
			c) DeepImpact                                & 0.326                             &    0.696    &         0.472          &    0.650    &         0.426          &    n.a.    &            235            \\
			d) uniCOIL                                   & 0.351                             &    0.693    &         0.476          &   0.666     &          0.445          &    240     &            276           \\ \midrule
			\textbf{(ii) Dense Retrievers}            &                                   &             &                        &             &                        &         &                            \\
			e) RepBERT                                   & 0.304                             &    0.610    &         0.331          &    0.662    &         0.370          &  152   &      1,633            \\
			f) ANCE                                      & 0.330                             &    0.645    &         0.361          &    0.642    &         0.405          &  152   &           1,633            \\ \midrule
			\textbf{(iii) Bert-based Re-rankers}      &                                   &             &                        &             &                        &         &                            \\
			g) EPIC+BM25-top100                             & 0.274                             &    0.608    &         0.411          &    0.573    &         0.349          &  96 &         113       \\
			h) EPIC+d2q-top15            & 0.303                             &    0.691    &         0.473          &    0.628    &         0.406          &  101  &          116      \\
			i) BERT-base+BM25-top1000                     &0.350                            &    0.706    &         0.483          &    0.686    &         0.454          & $3,815$ &            n.a.         \\
			j) BERT-large+BM25-top1000                    & 0.370                             &    0.738    &         0.506          &     0.705     &          0.493           & $11,594$ &            n.a.      \\ \midrule
			\textbf{(iv) Tokenizer-based Re-rankers} &                                   &             &                        &             &                        &         &                            \\
			k) TILDE+BM25-top1000                              & 0.269                             &    0.579    &         0.406          &    0.620    &         0.406          &   n.a.   &         76.6      \\
			l) TILDE+d2q-top10              & 0.285                             &    0.650    &         0.467          &    0.624    &         0.417          &   n.a.   &        75.3    \\
			TILDEv2 (ours)+BM25-top1000                        & $0.333^{abeghjkl} $                          &    $0.676^a$    &         $0.448^{ae}$          &    $0.659^{ag}$     &          $0.433^{agj}$          &   n.a.   &         80.8        \\
			TILDEv2 (ours)+d2q-top100       & $0.341^{abeghjkl} $                             &    $0.703^{agk}$    &         $0.498^{abefgk}$          &    $0.669^{ag}$     &          $0.449^{ag}$         &   n.a.  &         76.4       \\ \bottomrule
		\end{tabular}
	\label{table:results}
\end{table*}

\subsection{RQ1: Effectiveness and efficiency of TILDEv2 vs. TILDE}
Table~\ref{table:results} reports the results obtained with respect to effectiveness (MRR@10, nDCG@10, MAP) and efficiency (query latency) across the three studied datasets. For the baseline re-rankers, we use the best re-ranking cut-off reported in the respective original papers. For TILDEv2, we tune the cut-off on a subset of dev queries from MS MARCO, and use docT5query as the passage expansion method for fair comparison with DeepImpact and uniCOIL, which also use docT5query for passage expansion.

We first start by comparing the original TILDE and our TILDEv2 (block iv in Table~\ref{table:results}), thus answering RQ1: which matching mechanism is more effective. Both methods in fact only use the BERT tokenizer to encode the query, but rely on different matching mechanisms. The results indicate that the contextualized exact term matching employed in TILDEv2 leads to higher effectiveness than the reliance on query likelihood of the original TILDE; and this is regardless of the first stage of retrieval (BM25 vs docT5query), though docT5query leads to better results than BM25.
These improvements are especially significant for the MS MARCO dataset, with 24\% when re-ranking BM25 and 20\% when re-ranking docT5query. 

When considering query latency, we observe that both methods only require less than 100 ms to generate the final ranking. In addition, TILDEv2 is only 4.2 ms slower than the original TILDE when re-ranking BM25, and 1.1 ms slower when re-ranking docT5query. We note that when tuning the rank cut-off on a random sample of dev queries (see above), TILDEv2 was found to be most effective when re-ranking the top 100 passages, while the original TILDE used the top 10 passages: that is, although TILDEv2 re-ranking takes a handful of milliseconds more than TILDE, it does re-rank more passages. Furthermore, an additional reduction in runtime could be achieved by optimizing the index structure used by TILDEv2. 
 
 In summary, in answer to RQ1, we conclude that the contextualized exact term matching of TILDEv2 leads to better effectiveness (significantly on MS MARCO) than the query likelihood matching used by the original TILDE, at no or minor expense of query latency.

\subsection{RQ2: Effectiveness and efficiency of TILDEv2 vs. baselines}

We now compare TILDEv2 with other baselines in terms of effectiveness (RQ2). For MS MARCO, TILDEv2 outperforms, significantly, most comparison methods, although it is outperformed by uniCOIL and the BERT-base/large re-rankers. Differences between TILDEv2 and uniCOIL and BERT-base are not statistically significant; while those with BERT-large are.
Similar results are observed for the two TREC DL datasets, with the difference that for these datasets TILDEv2 displays better effectiveness than uniCOIL when re-ranking docT5query (no statistical significance).

The biggest advantage of TILDEv2 is however the low query latency. TILDEv2 does not require GPUs and only adds a couple of milliseconds on top of the first stage retrieval (BM25 or docT5query), overall resulting in a query latency of $\le80$ ms on CPU. The other neural methods, instead, either require GPUs to achieve acceptable query latency (and in any case higher than that of TILDEv2) or, if they can feasibly be used on CPUs, display higher query latency. 
Table~\ref{table:query_latency} further details the breakdown of query latency for TILDEv2 and some strong baselines in the CPU environment. For uniCOIL, ANCE and EPIC, even just the query processing amounts to about 50\% of the total latency of TILDE and TILDEv2. For DRs (ANCE), the retrieval time is much higher on CPU than GPU. The latency of uniCOIL is acceptable on CPU; yet, it is more than 4 times higher than that of the original TILDE and of TILDEv2. EPIC, TILDE and TILDEv2 have comparable query latency, but EPIC and TILDE are much less effective than TILDEv2 (as shown in Table~\ref{table:results}).

In summary, in answer to RQ2, TILDEv2's effectiveness is on par with or better than the considered baselines, while achieving much higher efficiency. 

\subsection{RQ3: Effectiveness-efficiency trade-off}

In this section, we investigate the effectiveness-efficiency trade-off of TILDEv2 (RQ3). This trade-off appears in many second-stage re-ranking methods with respect to the rank cut-off used for the re-ranking: higher cut-offs translate to more passages needing to be evaluated, and thus higher query latency. For this analysis, we compare our TILDEv2, which ran on a CPU environment, with the BERT-large re-ranker, ran on a GPU environment, on the task of re-ranking the top $k$ results from BM25. The BERT-large re-ranker is the strongest re-ranker baseline we considered. 

\begin{figure}
	\includegraphics[width=\linewidth]{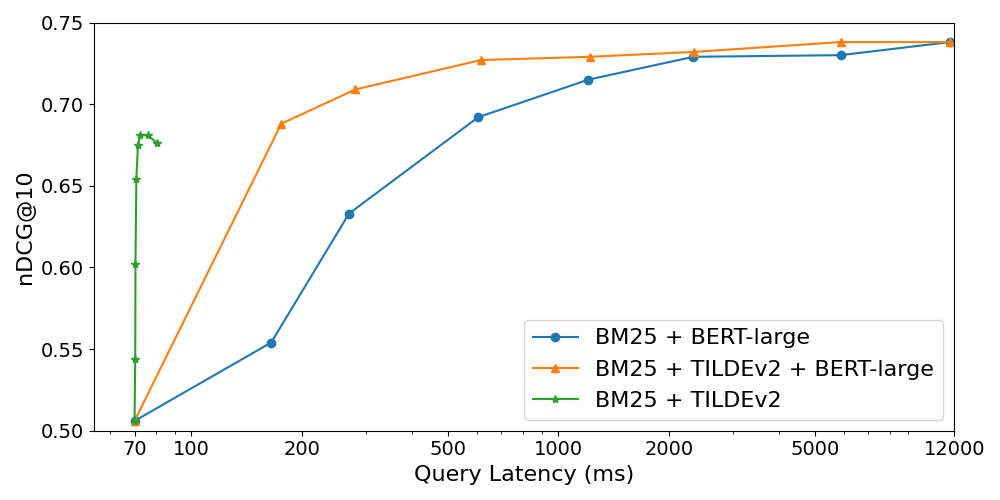}
	\caption{Query latency analysis vs. nDCG@10 on TREC DL2019. Points from left to right are re-ranking cut-offs $k=$ 0 (no re-rank), 10, 20, 50, 100, 200, 500 and 1,000. \vspace{-12pt}}
	\label{fig:rerank}
\end{figure}

In  Figure~\ref{fig:rerank} we report the analysis of this effectiveness-efficiency trade-off  in terms of nDCG@10 vs. query latency measured on TREC DL 2019 (similar results on other datasets), exploring the cut-offs $k=\{0, 10, 20 ,50, 100, 200, 500, 1000\}$ ($k=0$: no re-ranking).

The BERT-large re-ranker (blue line) achieves higher effectiveness when considering deeper cut-offs, as reported before by others~\cite{lin2020pretrained,wang2021bert}. This is problematic because the BERT-large re-ranker is very inefficient when evaluating a passage (requiring on average 12ms): thus the cut-off conditions which make the BERT-large re-ranker highly effective are the same that make it largely inefficient -- it takes around 12,000 ms to re-rank the top 1,000 passages.

In contrast, TILDEv2 (green line) only needs a few milliseconds to re-rank the top 1,000 passages, and re-ranking less than that leads to negligible time-saving. What is more, in less than 80ms TILDEv2 can achieve the same effectiveness that the BERT-large re-ranker achieves after spending more than 500 milliseconds (cut-off k=50).

Finally, we also consider a three-stage re-ranking system (orange line), where the top 1,000 BM25 results are first re-ranked using TILDEv2 and that ranking is further processed using the BERT-large re-ranker on the top $k$ passages. The use on the intermediate re-ranking step with TILDEv2 adds very little to the overall latency: just 10.8 ms. 
As shown in Figure~\ref{fig:rerank}, the use of the intermediate TILDEv2 step allows the final BERT re-ranker to produce highly effective results using smaller rank cut-offs (i.e. $10 \le k \le 50$), thus overall producing substantial savings in query latency compared to when TILDEv2 is not added to the pipeline (blue line).
For example, to reach the same level of effectiveness reached by the BERT-large re-ranker with $k=50$ (latency 605 ms), the three-stage pipeline with TILDEv2 only requires the BERT-large re-ranker to re-rank the top 10 TILDEv2 results (latency 175 ms), with a saving of 430 ms. Similarly, the effectiveness reached by the three-stage pipeline with $k=50$ is not statistically significantly different from that reached by the BERT-large re-ranker alone with $k={200, 500, 1000}$.
 
 In summary, in answer to RQ3, we find that the rank cut-off $k$ controls the efficiency vs. effectiveness trade-off for the BERT-large re-ranker: low values of $k$ result in lower latency and lower effectiveness, while larger $k$ values yield more effective results but substantially higher latencies. The rank cut-off $k$, however, while impacting TILDEv2's effectiveness (higher values lead to higher effectiveness), barely has any effect on query latency. Furthermore, we find that injecting TILDEv2 into the BERT-large re-ranker pipeline allows to reach the best effectiveness produced by the BERT-large re-ranker alone, for a fraction of the latency.

%

\begin{table}[t]
	\caption {Detailed query latency in CPU environment, in milliseconds. Each method is set to use the parameters that best optimise MRR@10 on MS MARCO. ANCE and uniCOIl do not perform re-ranking. \vspace{-10pt}} 
		\begin{tabular}{ l |c |>{\small}c |>{\small}c |>{\small}c}
			\toprule			
			\bf Methods & \thead{Query \\ process}&  Retrieval&  Re-rank & Total \\
			\toprule  
			uniCOIL       & 46     & 230   & n.a. & 276 \\
			ANCE           & 63  & 1,570  & n.a. & 1,633   \\
			EPIC+d2q-top15 & 40 & 75 & 1 & 116  \\
			TILDE+d2q-top10  & 0.1 & 75 & 0.2 & 75.3\\
			TILDEv2+BM25-top1000 & 0.1 & 70 & 10.7 & 80.8 \\
			TILDEv2+d2q-top100  & 0.1 & 75 & 1.3 & 76.4 \\
			
			\bottomrule
		\end{tabular}
	\label{table:query_latency}
\end{table}

\begin{table*}[t]
	\caption {Impact of different expansion methods. The index size estimation of TILDEv2 is included BM25 index size and contextualized term weight index size. The expansion cost is estimated base on Google cloud service. }
	\label{table:expansion}
	\begin{tabular}{l|c|c|c|c|c|c|c|c|c|c}
		\toprule
		& \multicolumn{2}{c|}{\bf no Expansion} & \multicolumn{2}{c|}{\bf docT5query} & \multicolumn{2}{c|}{\bf TILDE, $m = 128$}  & \multicolumn{2}{c|}{\bf TILDE, $m = 150$}  & \multicolumn{2}{c}{\bf TILDE, $m = 200$}  \\ \hline
		& uniCOIL     & TILDEv2       & uniCOIL        & TILDEv2        & uniCOIL         & TILDEv2       & uniCOIL         & TILDEv2       & uniCOIL         & TILDEv2       \\ \hline
		MS MARCO, MRR@10       & 0.319       & 0.299         & 0.351          & 0.333          & 0.343           & 0.326         & 0.346           & 0.327         & 0.349           & 0.330         \\ 
		DL2019, nDCG@10  & 0.653       & 0.613         & 0.693          & 0.676          & 0.682           & 0.680         & 0.690           & 0.679         & 0.707           & 0.670         \\
		DL2019, MAP      & 0.418       & 0.417         & 0.476          & 0.448          & 0.464           & 0.457         & 0.470           & 0.452         & 0.474           & 0.447         \\  \Xhline{2.5\arrayrulewidth}
		Index size      & 4.3G        & 4.8G    & 6.0G           & 5.2G           & 5.6G            & 5.2G          & 6.7G            & 5.6G          & 9.81G           & 6.9G          \\ \hline
		Avg added token & \multicolumn{2}{c|}{0}      & \multicolumn{2}{c|}{19.0}       & \multicolumn{2}{c|}{13.0}       & \multicolumn{2}{c|}{25.2}       & \multicolumn{2}{c}{61.6}       \\ \hline
		Expansion cost  & \multicolumn{2}{c|}{0}      & \multicolumn{2}{c|}{320 hours/768\$}  & \multicolumn{2}{c|}{7.22 hours/5.34\$} & \multicolumn{2}{c|}{7.25 hours/5.37\$} & \multicolumn{2}{c}{7.33 hours/5.42\$} \\ \bottomrule
	\end{tabular}
\end{table*}

\begin{table*}[t]
	\caption {Tokens generated by docT5query and TILDE for the first passage in the MS MARCO dataset (pid=0). }
	\label{table:case}

		\begin{tabular}{  p{4.8cm} | p{2.5cm} |p{2.8cm} | p{6.2cm}  }
			\toprule
			\bf Original passage& \bf  docT5query & \bf  TILDE, m=128 & \bf  TILDE, m=200 \\
			\hline
			the presence of communication amid scientific minds was equally important to the success of the manhattan project as scientific intellect was. the only cloud hanging over the impressive achievement of the atomic researchers and engineers is what their success truly meant; hundreds of thousands of innocent lives obliterated. 
			&      amongst scientists why? about so a importance purpose how significant in for believe who did     
			&      importance purpose quiz scientists bomb genius development solving significance successful intelligence solve effect objective research  accomplish brains progress scientist     
			&    ... future impact strategic develop necessary ni role involved developing needed theory significant technology achievements accomplished science achieve intellectual new breakthrough help keypower effects effort human work engineer concept invention idea problem process ability communicate developed would affect solved decision use deal society reason effective franklin problems great goals opportunity secret considered      \\
			\bottomrule   
		\end{tabular}

\end{table*}

\subsection{RQ4: Impact of passage expansion}\label{sec:expansion_impact}
In this section, we investigate the impact of different passage expansion methods (RQ4), and in particular the docT5query and the proposed TILDE for passage expansion, with respect to the cost of the passage expansion process (time and money), the quality of expanded terms and the impact on index size. To study these aspects, we consider passage expansion in the context of our TILDEv2 when re-ranking the top 1,000 passages from BM25, and of uniCOIL for full index retrieval~\cite{lin2021few}. Both methods require expanded passages at training and ranking; the use of uniCOIL allows us to verify the generalisability of the TILDE passage expansion method. The methods are tested with no expansion, docT5query expansion, and TILDE expansion. For docT5query we generate 40 expansion queries, as done in previous work; larger values are infeasible (very long generation time). For TILDE, we generate $m={128, 150, 200}$ expansion terms. Recall that not all expansion terms are added to a passage: only new expansion terms are added.

The results are reported in Table~\ref{table:expansion}. 
Any form of expansion improves over the not expanded results, showing that passage expansion is crucial for both TILDEv2 and uniCOIL. 
The docT5query method produces the most effective expansions for both methods on MS MARCO, although differences between docT5query and TILDE $m=200$ are marginal (0.349 vs. 0.351 for uniCOIL, 0.330 vs. 0.333 for TILDEv2) and not statistically significant ($p=0.848$ and $p=0.663$, respectively).
TILDE $m=200$ provides the most effective expansions for uniCOIL on TREC DL2019 (for nDCG@10), and TILDE $m=128$ provides the most effective expansions for TILDEv2. For MAP on this dataset, docT5query provides the most effective expansions for uniCOIL, while  TILDE $m=128$ provides the most effective ones for TILDEv2. Overall, the effectiveness achieved by the TILDE expansion method is on par with that of docT5query.


Next, we consider the size of the index produced by the methods (for TILDEv2, this is the size of the inverted index for BM25 and the term weight index, implemented using HashTable). When TILDE $m = 128$ is used, the index size produced by uniCOIL is smaller than when using docT5query, while the one produced by TILDEv2 is the same as if docT5query was used for expansion. When $m$ increases, the index size increases for both uniCOIL and TILDEv2: this is expected as an increase in $m$ will make TILDE produce more tokens that are added to the passages. This can be seen by observing the average added token row in Table~\ref{table:expansion}. However, two observations can be made: (1)
 the index produced by TILDEv2 is always smaller than that produced by uniCOIL, despite the same number of tokens being added, and (2) the size of the TILDEv2 index is two orders of magnitude smaller than the size of the original TILDE index (not reported in the table), which is $\approx$500Gib~\cite{tilde2021zhuang}.


We then consider the cost of executing the two passage expansion processes on the whole MS MARCO passage collection. The TILDE expansion process is two orders of magnitude faster to run than the docT5query (320 hours vs. $\approx$7 hours). In addition, the TILDE expansion process can be run on a preemptible GPU environment, which is much cheaper than the preemptible TPU environment used by Nogueira et al.~\cite{nogueira2019doc2query} to run the docT5query expansion\footnote{Using GPUs would result in a larger runtime.} (costs estimated based on the Google cloud service). Thus, using our TILDE expansion method is not just faster than docT5query, but also cheaper. These advantages become even more obvious when considering expanding larger collections such as the MS MARCO v2 (138.3 million passages): it would take docT5query $\approx$ 5,000 hours and \$12,000, while only requiring TILDE $\approx$114 hours and \$85.


Table~\ref{table:case} reports the tokens added by different expansion methods for an example passage.  The expanded tokens from docT5query contain symbols such as the question mark, and common stopwords, e.g., `a', `for'. On the other hand, TILDE removes stopwords and symbols before appending to the passage. Interestingly, both docT5query and TILDE add similar tokens such as `scientists', `importance', and `purpose'.  In general, the added tokens from both methods are on-topic. 

In summary, in answer to RQ4, our TILDE passage expansion method achieves similar effectiveness as the state-of-the-art docT5query, but it requires far fewer computational resources.

\section{Conclusion} \label{conclusion}
We proposed the TILDEv2 model for passage re-ranking, which builds on top of the recently proposed TILDE by integrating the best-of-breed from recent advances in neural retrieval. We further proposed a novel use of the original TILDE as an effective and efficient passage expansion technique. Our TILDEv2 aims to solve some of the drawbacks of the original TILDE model (effectiveness, large indexes) by integrating the contextualized exact term matching approach. While, our passage expansion technique aims to address the scalability issues of current methods for passage expansion.


The empirical results show that  TILDEv2 significantly improves the effectiveness of the original TILDE and largely reduces its index size, while maintaining its efficiency and without resorting to expensive computational environments (TILDEv2 runs on CPU).
We also find that the proposed TILDE-based passage expansion method delivers computational cost savings of up to 98\% compared to other passage expansion methods, while experiencing effectiveness drops of less than 1\% (and improving effectiveness in certain settings). The proposed passage expansion method can be used not just with TILDEv2: we show its performance generalises to other methods such as uniCOIL. 
These results make TILDEv2 a production-ready method of great appeal in search settings that require low query latency and have limited computation resources available. The code that implements TILDEv2 and that can be used to reproduce the results in this paper is available at \url{https://github.com/ielab/TILDE/tree/main/TILDEv2}.




\bibliographystyle{ACM-Reference-Format}
\bibliography{WSDM2022-TILDEv2}

\appendix

\end{document}